\documentclass{article}
\usepackage[utf8]{inputenc}
\usepackage{charter}
\usepackage{fullpage}
\usepackage{hyperref}
\usepackage{amssymb}
\usepackage{graphicx}
\usepackage{lipsum}
\usepackage{hyperref}
\usepackage{upgreek}
\usepackage{comment}
\hypersetup{
colorlinks = true,
            linkcolor = black,
            urlcolor  = blue,
            citecolor = blue}
\usepackage[colorinlistoftodos]{todonotes}
\setuptodonotes{inline}
\usepackage[hybrid]{markdown}
\usepackage{array,multirow,graphicx}
\usepackage[thinlines]{easytable}
\usepackage{makecell}
\usepackage[margin=0.99in]{geometry}
\usepackage{siunitx}
\usepackage{placeins}

\newcommand\snowmass{
\begin{center}
  \rule[-0.2in]{\hsize}{0.01in}\\
  \rule{\hsize}{0.01in}\\
  \vskip 0.1in
  Submitted to the Proceedings of the US Community Study\\ 
  on the Future of Particle Physics (Snowmass 2021)\\
 
  \rule{\hsize}{0.01in}\\
  \rule[+0.2in]{\hsize}{0.01in}\\[-2em]
\end{center}
}

\usepackage[firstpage=true]{background}
\backgroundsetup{contents={\parbox{6.5in}{\snowmass}}, scale=1,placement=top,opacity=1,color=black,position={3.25in,1.2in}}

\usepackage{fancyhdr}
\fancypagestyle{plain}{%
  \fancyhf{}%
  \fancyhead[C]{}
  \fancyfoot[C]{\thepage}
}

\title{Supporting Capabilities For Underground Facilities}
\date{}

\usepackage{authblk}

\author[1]{Alvine Kamaha\thanks{corresponding author:akamaha@physics.ucla.edu}}

\author[2]{Brianna Mount}

\author[3]{Richard Schnee}

\affil[1]{University of Califonia, Los Angeles, Department of Physics \& Astronomy, Los Angeles, CA 90095-1547}

\affil[2]{Black Hills State University, School of Natural Sciences, Spearfish, SD 57799-0002, USA}

\affil[3]{South Dakota School of Mines and Technology, Rapid City, SD 57701-3901, USA}
\begin{document}

\maketitle
\begin{abstract}
The 2021 particle physics community study, known as “Snowmass 2021”, has brought together particle physicists around the world to create a unified vision for the field over the next decade. One of the areas of focus is the Underground Facilities (UF) frontier, which addresses underground infrastructure and the scientific programs and goals of underground-based experiments. To this effect, the UF Supporting Capabilities topical group created two surveys for the community to identify potential gaps between the supporting capabilities of facilities and those needed by current and future experiments. Capabilities surveyed are discussed in this report and include underground cleanroom space size and specifications, radon-reduced space needs and availability, the assay need and other underground space needs as well timeline for future experiments.  Results indicate that future, larger experiments will increasingly require underground assembly in larger, cleaner cleanrooms, often with better radon-reduction systems and increased monitoring capability for ambient contaminants.  Most assay needs may be met by existing worldwide capabilities with organized cooperation between facilities and experiments. Improved assay sensitivity is needed for assays of bulk and surface radioactivity for some materials for some experiments, and would be highly beneficial for radon emanation.
\end{abstract}
\tableofcontents

\section{Introduction}
\label{Intro}
Underground experiments require significant supporting capabilities, including above-ground and underground cleanrooms, radon-reduction systems, and low-background assay systems. These capabilities are required to create and maintain low-radioactive environments for the operation of radiation-sensitive experiments such as those described in other Underground Facility reports 
for 
neutrino physics and dark matter. 
To assess the 
needed supporting capabilities for future experiments,
a survey was sent to all current and 
planned underground experiments with SNOWMASS white papers. Concurrently, a survey was sent to all current and planned underground facilities. Tables~\ref{ExpSurvey_respondent} and~\ref{FacSurvey_respondent} list all survey respondents.  Based in great part on the responses, Sections~\ref{SupportFacilities}--\ref{SupportOther} below describe facilities’ supporting capabilities and the needs of future experiments.

\begin{table}[ht]
\caption{Survey respondents: List of Experiments} 
\begin{center}
\begin{tabular}{|l|l|l|}
\hline
\multicolumn{3}{|c|}{\textbf{Current Experiments}}   \\
\hline
CANDLES~\cite{CANDLES2022PANIC}     & DM-Ice~\cite{DM-ICE2017}              & NEXT-100~\cite{NEXT2015} \\
CDEX~\cite{CDEX2022}                & Hyper-Kamiokande~\cite{Hyper-Kamiokandesnowmass}  & PandaX~\cite{PandaX4T2022backgrounds,PANDAX4T2022}      \\
COSINE-100~\cite{COSINE-100:2021}   & KamLAND-Zen~\cite{KamLANDZen2022}     & Super-Kamiokande~\cite{Super-Kamiokande2021}                \\
CUPID~\cite{CUPID-MO2022,CUPID2022} & Majorana Demonstrator~\cite{MJD2022} &   SNO+~\cite{albanese2021sno+}\\
\hline
\multicolumn{3}{|c|}{\textbf{Planned Experiments}}   \\
\hline
Argo~\cite{DarkSide2022HEP2021}     & DARWIN~\cite{DARWIN2021,XLZD2022}     &  NEXT w/Ba-Tagging~\cite{NEXT2020,NEXTBaTag2019DPF}        \\
COSINE-200~\cite{COSINE2020}        & Kton Xe TPC for 0$\nu\beta\beta$~\cite{ktonBetaBeta}  & NuDot~\cite{NuDOTloi,Gruszko_2019}          \\
CUPID-1T~\cite{CUPID1Tsnowmass}     & LEGEND~\cite{LEGENDPCDR2021}          & PIRE-GEMADARC~\cite{UndergroundGe,PIRE-GEMADARCloi}              \\
CYGNUS~\cite{CYGNUS1000concept,CYGNUSsnowmass}  & nEXO~\cite{nEXO2021}      & SBC~\cite{SBC2021,SBC2022}      \\
DarkSide-20k~\cite{DarkSide2022HEP2021}    & NEXT-CRAB~\cite{NEXT2020,NEXTBaTag2019DPF} & Snowball~\cite{snowball}                      \\
DarkSide-LowMass~\cite{DarkSide2022HEP2021} & NEXT-HD~\cite{NEXT2020}       & 50-ton bubble chamber~\cite{PICOloi} \\
\multicolumn{3}{|l|}{A possible neutrinoless-double beta-decay extension to DUNE~\cite{DUNEbetabeta}}  \\
\hline
\end{tabular}
\label{ExpSurvey_respondent}
\end{center}
\end{table}
\begin{table}[ht]
\caption{Survey respondents: List of Facilities} 
\begin{center}
\begin{tabular}{|l|l|}
\hline
Berkeley Low Background Counting Facility, U.S.~\cite{BLBC2015}     &  Boulby, UK~\cite{Boulby2012,BoulbyGammas,BoulbyNeutrons} \\
Canfranc, Spain~\cite{CanfrancRadon,CanfrancNeutrons,CanfrancRadonMitigation}    &  Gran Sasso, Italy~\cite{GranSasso,GranSassoNeutrons}    \\
JinPing, China~\cite{JinPing2021,JinPingNeutrons}                   &  Kamioka Observatory, Japan~\cite{Kamioka2012}  \\
KURF, VA, U.S. (not available due to COVID)~\cite{Kimballtonloi,KimballtonGe,KimballtonLRT2010} & LAFARA, French Pyrénées~\cite{LAFARA}  \\
LLNL Nuclear Counting Facility, U.S.                                & Modane, France~\cite{Modane2012,Modane2020backgrounds,ModaneNeutrons}  \\
Pacific Northwest National Laboratory, U.S.~\cite{PNNLCASCADES}     & SNOLAB, Canada~\cite{SNOLAB2020} \\
SURF, SD, U.S.~\cite{SURFwp}                                        & Y2L / Yemilab, Korea~\cite{Yemilab2021,Yemilab2020,YangYangneutrons2021}   \\
 U. Alberta, Canada~\cite{LRT2010HallinRadon}                       & SD Mines, SD, U.S.~\cite{LRT2015streetVSA,LRT2017streetVSA} \\
\hline
\end{tabular}
\label{FacSurvey_respondent}
\end{center}
\end{table}

\section{Facilities for Low-Radioactivity Fabrication and Assembly}
\label{SupportFacilities}
A general need for most underground experiments is space for low-radioactivity fabrication and assembly.  Cleanrooms (as described in Sect.~\ref{SupportCleanrooms}) and radon-reduced air environments (as described in Sect.~\ref{RadonReducedSpaces}) are important supporting facilities to mitigate exposure to ambient background sources. 

\subsection{Cleanroom Capabilities} 
\label{SupportCleanrooms}


Dust on or in sensitive detectors can compromise their operation (e.g.\ by causing electrical shorts or sparking~\cite{lindley1970feed}) and increase their radioactive backgrounds since dust particulates may contain $^{238}$U, $^{232}$Th and $^{40}$K~\cite{akerib2020lux, akerib1802projected, lz_assay_2020, di2021direct, mount2017lzTDR}.  Dust may also emanate radon into detector active volumes after detector assembly~\cite{lz_assay_2020}.
It is therefore often critical to minimize exposure of detector materials to dust at all stages of storage, handling, and detector assembly. 
The higher level of mine dust 
in many underground spaces increases the level of contamination of the detector surfaces by these particulates compared to above ground if dedicated cleanroom spaces are not used. 

Detectors for underground experiments 
have often been
assembled in cleanroom laboratories above ground and then transported underground to finalize the assembly. As the need for bigger detectors arises for the future of these  experiments 
, larger underground clean areas will be needed for detector assembly, as transport of very large assembled detectors from the surface will become 
too difficult.  Furthermore, stricter background requirements will limit the amount of time materials can be at the surface, suffering cosmogenic activation.  Underground clean areas will also be increasingly needed for material storage, screening facilities, and detector development such as crystal growth for solid state detectors~\cite{WangGeCrystals2015, Ma_2018GeCosmogenicActivation}.

\begin{figure}[ht]
    \centering
    \includegraphics[width=0.7\textwidth]{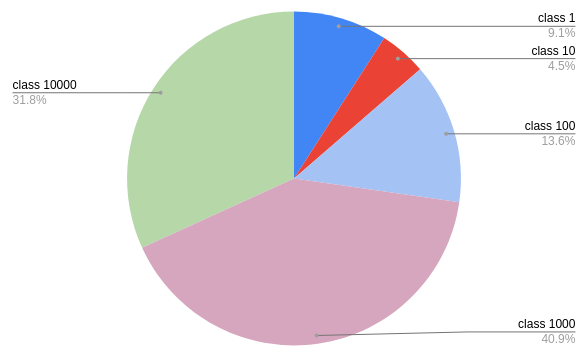}
    \caption{Cleanroom class requested by future UG experiments   
    \label{CleanroomPiechart}}
\end{figure}

The standard cleanroom ISO-6--7 (class 1000--10000) currently available in different facilities across the world is sufficient for many experiments but not for all experiments. Some experiments require improving 
these cleanrooms to ISO-5 (class 100) for further suppression against dust fallout onto the detector material surfaces during the assembly stage, as shown in Figure~\ref{CleanroomPiechart}.
Table~\ref{TableCleanooms} lists the cleanroom sizes and ISO classes available in underground and surface laboratories worldwide.

\begin{table}[ht]
\caption{Cleanroom spaces for underground facilities} 
\begin{center}
\begin{tabular}{|lrrc|}
\hline
		            & Depth & CR Areas & CR ISO \\
Laboratory          & (mwe) & (m$^2$) & Class  \\
\hline
Boulby, UK.         & 2805 &800   &ISO 7  \\
Canfranc, Spain~\cite{CanfrancRadon,CanfrancNeutrons} &  2400    &   70, 30       &  ISO 5-6   \\
Gran Sasso, Italy   & 3100 & 13      &    ISO 7     \\
Gran Sasso, Italy   & 3100 & 86, 32      &   ISO 6     \\
Gran Sasso, Italy   & 0    & 325       &ISO 6  \\
Gran Sasso, Italy   & 0    & 62        &  (in progress)\\
SNOLAB, Canada      & 5890 & 4924 & ISO 6-7 \\
SNOLAB, Canada      & 5890 &  3159 & ISO 6-7 \\
SURF, SD, U.S.      & 0    & 37         &ISO 6      \\
SURF, SD, U.S.      & 0    & 55         &ISO 5-6      \\
SURF, SD, U.S.      & 4300 & 120, 56, 55, 41         &ISO 5-6      \\
SURF, SD, U.S.      & 4300  & 52, 18.3         &  ISO 6-7 \\
SURF, SD, U.S.      & 4300  & 286, 125, 38, 34         &  ISO 7 \\
SURF, SD, U.S.      & 4300  & 90         &  ISO 8 \\
Y2L, Korea                 & 1750 & 46, 46    &  ISO 7  \\
Yemilab (under construction), Korea   & 2800 & 23   & ISO 5 \\
Yemilab (under construction), Korea   & 2800 & 80, 20   & ISO 7 \\
Kamioka Observatory, Japan & 2700  &  66  &   Not relayed \\ 
PNNL, U.S.  & 38 & 5$\times$19-60 & ISO 6-7 \\
\hline
\end{tabular}
\end{center}
\label{TableCleanooms}
\vspace{-0.3cm}
\end{table}

For these future detectors' development and assembly, multiple-sites monitoring of the dust concentration within the cleanrooms as well as the dust fallout rate over time is also recommended. 
Particulate detectors should be distributed in strategic areas to sample the air within the room over time with prompt feedback. Collection vials or Witness plates should also be distributed in these areas to be measured with ICP-MS or optical and/or x-ray fluorescence microscopy to enable an accurate modelling and tracking of the dust content within the room and its deposition onto the detector materials (which can be confirmed later with tape-lift measurements). 
The lowest requirements on dust fallout rate is at the level of 100\,ng/cm${^2}$ over the duration of experiment assembly for inner detector surfaces with a requirement of $\sim$10${^{-17}}$\,g (U,Th) /cm${^2}$ on U and Th from dust. These requirements are 
modestly lower than the sensitivity of the current microscopy techniques for dust deposition but may be met for long-lived isotopes using ICP-MS~\cite{diVacri2021dust}.


\subsection{Radon-reduced Cleanrooms and Other Spaces}
\label{RadonReducedSpaces}


Radon-daughter plate-out onto detector surfaces during storage, handling, or detector assembly provides additional long-lived radioactive contamination for underground experiments.  Contamination with $^{210}$Pb ($t_{1/2} =$ 22.3 year) contributes to experimental backgrounds long after the initial plate-out via its beta decay~\cite{SCDMSsensitivity,leung2005borexino,xenonEMbackgrounds2011,lux2014backgrounds}, alpha decay~\cite{lux2013nim,lux2015RnBackgrounds,mount2017lzTDR,Breunner2021PTFERnRemoval} and recoiling daughters~\cite{SCDMSsensitivity,cresst2012,deap2011surface,deap2015Rn,coupp2012,mount2017lzTDR,Xu2017Ar206Pb}.
Due to nuclear recoil momentum, decay daughters are generally embedded 
tens of nm into the detector material surfaces after the initial parent depositions.
The contaminants are therefore not easily removed with remedial cleaning after the assembly is complete. Techniques such as acid etching or electropolishing may be performed in some cases with relatively good efficiencies at removing some of the implanted radon daughters ($^{210}$Pb,  $^{210}$Bi, $^{210}$Po)~\cite{schnee2013EPRnRemoval,zuzel2015RnRemoval,Guiseppe2018Po210removal,CDMSsidewallEtch2020NIM,Breunner2021PTFERnRemoval,BUNKER2020163870}. The best approach remains 
mitigation against the deposition of 
radon daughters onto the detector material surfaces.

The air in underground laboratories typically has a high radon concentration ($\sim$100\,Bq/m$^3$)~\cite{JinPingRadon,KamiokaRadon}, although some underground sites (such as Boulby and KURF) have low radon concentrations similar to outdoors ($\sim$5\,Bq/m$^3$) throughout their entire facilities. 
Many experiments require cleanroom areas for detector fabrication and assembly with radon concentrations below that of outside air.
Larger future detectors requiring lower levels of radon-daughter plate-out will also necessitate larger cleanrooms underground with even lower radon concentrations.  
Table~\ref{TableRadonFacilities} lists the current low-radon cleanrooms worldwide  along with additional spaces with radon concentrations reduced to lower than outside air. In general, these facilities have been built to meet the needs of specific near-term experiments.  Future experiments described above, 
such as liquid noble detectors, tend to need reduced-radon cleanrooms with areas 100--200 m$^2$, while several next-generation experiments (such as DarkSide-LowMass and future phases of NEXT) require lower radon concentrations (1--5\,mBq/m$^3$) than are currently available.  These lowest radon concentrations desired are at, but not beyond, the capabilities of the most sensitive radon monitors so far produced.



\begin{table}[ht]
\caption{Radon-reduced spaces for underground facilities} 
\begin{center}
\begin{tabular}{|lrrlrc|}
\hline
		            & Depth & CR Area & CR ISO & Rn Concentration   & Other \\
Laboratory          & (mwe) & \multicolumn{1}{c}{(m$^2$)} & Class  & \multicolumn{1}{c}{(mBq/m$^3$)}   & Areas \\
\hline
Canfranc, Spain~\cite{CanfrancRadonMitigation} & 2400 & 70~~~ & ISO 5-6 & $<$5~~  & 1 mBq/m$^3$ to experiments \\
Gran Sasso, Italy   & 3100 & 13~~~       & ISO 7  & 10~~     & \\
Gran Sasso, Italy   & 3100 & 86~~~       & ISO 6  & 50~~     & \\
Gran Sasso, Italy   & 3100 & 32~~~       & ISO 6  & 50~~     & \\
Gran Sasso, Italy   & 0    & 325~~~      & ISO 6  & (in progress)~~  & \\
Gran Sasso, Italy   & 0    & 62~~~       & ISO 6  & (in progress)~~   & \\
Kamioka Obs., Japan & 2700  & & & & 50 mBq/m$^3$ to SuperK tank \\
Modane, France~\cite{ModaneRadonMitigation}      & 4800 & 16~~~       &        & (planned)~~      & 15 mBq/m$^3$ to experiments \\
SNOLAB, Canada      & 5890 &          & ISO 6  & (in progress)~~    & \\
SURF, SD, U.S.      & 4300 & 45~~~       & ISO 7  & 100~~            & \\
SURF, SD, U.S.      & 0    & 55~~~       & ISO 5-6 & 500~~            & \\
Y2L                 & 1750 & 46~~~       & ISO 7  & 1000~~            & HPGe array room \\
Yemilab (planned)~\cite{Yemilab2020}        & 2800 & 23~~~       & ISO 5 & planned~~          & planned         \\
Yemilab (planned)~\cite{Yemilab2020}        & 2800 & 80~~~       & ISO 7 & planned~~          & planned         \\
U. Alberta, Canada~\cite{LRT2010HallinRadon}  & 0    & 100~~~      & ISO 5  & 100~~                 & \\              
SD Mines, U.S.~\cite{LRT2015streetVSA,LRT2017streetVSA}      & 0    & 15~~~       & ISO 5-6 & 20~~            & \\
\hline
\end{tabular}
\end{center}
\label{TableRadonFacilities}
\vspace{-0.2cm}
\end{table}


Because the ultimate goal of reduced-radon cleanrooms is to 
ensure a low level of radon-daughter plate-out onto detector surfaces is not exceeded,
monitoring of the radon daughter plate-out is also needed in many cases (especially since such plate-out rates depend not only on the radon concentration but also on the material charge and geometry). Such monitoring is typically achieved through a distribution of witness plates measured with low-background alpha detectors.  Desired sensitivities for many experiments are lower than 0.1\,mBq/m$^2$ activity of $^{210}$Po during a full construction period, implying that monitoring that can provide direct short-term feedback of use must be modestly better than the best sensitivity currently available.~\cite{xinranliuLRT2019XIA,BUNKER2020163870}

Some experiments require lower radon concentrations in the air surrounding their detectors (often in gaps within shielding layers).  Modane supplies air with a concentration of 15\,mBq/m$^3$ to its experiments, while Canfranc supplies 220\,m$^3$/hr air with 1 mBq/m$^3$~\cite{CanfrancRadon}.  Y2L provides purge gas with a concentration of 1\,Bq/m$^3$ to its HPGe detectors.  Several experiments use liquid nitrogen boil-off as described above.

\section{Assay needs}
\label{SupportAssays}

Underground experiments including dark matter searches and neutrinoless double beta decay experiments continue to require extreme detector radiopurity. Of particular interest are the primordial radionuclides, $^{40}$K, $^{232}$Th, and $^{238}$U which are present in most raw materials. For each of these experiments, materials are carefully screened and selected to comprise the detectors and their shielding. Once materials are selected, accurate and precise characterization is an important component in the modeling and analysis of their data. A complementary suite of assay capabilities, including High Purity Germanium (HPGe) Gamma-Ray Spectroscopy, Inductively Coupled Plasma Mass Spectrometry (ICP-MS), alpha screening, and radon emanation is required to determine which radionuclides are present in a material and at what levels, especially since decay chains are often not in secular equilibrium.~\cite{lz_assay_2020, PANDAX4Tscreening2022}. 

The surveyed current and planned experiments relayed a variety of needed sensitivities for sample assays, 
with most next-generation experiments aiming for $\sim$100\,nBq/kg assay capability for inner detector materials. However, KAMLAND-ZEN related their requirement of achieving on the order of 1\,nBq/kg. 

\subsection{High-Purity Germanium Gamma-Ray Spectroscopy} 
\label{supportHPGe}

Gamma-ray spectroscopy using HPGe detectors has historically been the workhorse of low-background efforts. These detectors are located in numerous underground lab around the world. Low-background counting of gamma rays to determine the radionuclides embedded within materials is sensitive down to \SI{10}{\micro\becquerel\per\kilogram} levels. Counting times for these detectors are routinely on the order of 1--2 weeks, 
with some up to a month in duration. Samples must be of sufficient mass to collect emission statistics but also must fit within the shielding of the detectors, which vary in size. HPGe is a non-destructive assay technique, so it can be used to assay final components.

For samples of smaller mass and activity, Neutron Activation Analysis (NAA) sometimes may be used~\cite{Tsang2021NAA}. Samples are first activated in a reactor, and then analyzed over a few weeks using HPGe detectors. This technique is effectively destructive to a low background sample as the sample is unusable after it is activated.  

\begin{table}
\caption{Current Low Background HPGe systems. 
Some sensitivities in our survey were not recorded.} 
\begin{center}
\begin{tabular}{|lrrc|}
\hline
		                                                &           &       & Sensitivity    \\
		                                                & Depth     & Number & [U], [Th]    \\
Facility                                                & (mwe)     & HPGe  &  (mBq/kg)    \\
\hline
Berkeley Low Background Counting Facility, U.S.~\cite{BLBFGe201347}         & 15        & 1~~   & 6 -- 24 \\
Boulby Underground Laboratory, UK~\cite{BoulbyGe}       & 2805      & 6~~   & $<0.1$ -- 1 \\
Canfranc, Spain~\cite{CanfrancGe2017127}                & 2400      & 7~~   & 0.1 -- 1   \\
China Jinping Underground Laboratory~\cite{JinPing2021}                    & 6720      & 3~~   & 1   \\
Gran Sasso, Italy~\cite{GranSassoGe,GatorGe}            & 3100      & 8~~   & 0.016 -- 15    \\
Kamioka Observatory, Japan~\cite{KamiokaRaGe2018,KamiokaRaGe2020}               & 2700      & 5~~   & $<1$ \\
LAFARA underground laboratory, French Pyrénées~\cite{LAFARA}                    & 220       & 5~~   & Not relayed \\
LLNL Nuclear Counting Facility, U.S.                    & 10        & 3~~   & Not relayed \\
Modane, France~\cite{ModaneHPGe2007,ModaneLowHPGe}      & 4800      & 2~~   & 0.4 -- 4      \\
Pacific Northwest National Laboratory, U.S.~\cite{PNNLGe,PNNLCASCADES}          & 38        & 14~~  & Not relayed \\
SNOLAB, Canada~\cite{SNOLABGe}                                          & 5890      & 5~~   & 0.04 -- 0.35     \\
SURF, SD, U.S.~\cite{BHUC2017130}                       & 4300      & 6~~   & 0.05 -- 0.7            \\
Vue-des-Alpes Laboratory, Switzerland~\cite{Vue-Des-Alpes,GeMSE2016,GeMSE2022}  & 620       & 1~~   & $<0.1$        \\
Y2L / Yemilab, Korea~\cite{YangYangGe2018o,Yemilab2020,KoreaHPGeArray14}        & 1750/2500 & 3~~   & 0.05 - 0.5 \\
SD Mines, U.S.                                          & 0         & 2~~   & 200 -- 2000             \\
\hline
\end{tabular}
\end{center}
\vspace{-0.2cm}
\label{TableHPGe}
\end{table}

As shown in Table~\ref{TableHPGe}, there are currently over 60 HPGe detectors serving underground experiments worldwide (and there are numerous HPGe detectors at additional underground laboratories not listed). If  
each detector counts a sample for two weeks and each detector requires four weeks of calibrations and background checks per year, the world-wide capability for ultra-low background counting is approximately 1,400 samples per year. 
Many experiments 
need on average 100 samples counted per year. However, limits of sensitivity for currently available HPGe may not reach the levels required by the most inner materials in the next generation of dark matter and neutrinoless double beta decay experiments. 
Current detector limits are on the order of \SI{10}{\micro\becquerel\per\kilogram}, about two orders of magnitude worse than needed for some materials. 
HPGe detectors with improved sensitivity (such as multiple-crystal detectors~\cite{KoreaHPGeArray14}), or other assay techniques with improved sensitivity, will be needed to provide assays for next-generation experiments.   
Furthermore, we cannot realize the full efficiency of having all world-wide detectors subscribed with the current model of each experiment ``owning” detectors. World-wide collaboration among low background counting labs is needed to fully realize the potential.

\subsection{Mass Spectrometry} 
\label{supportMS}

Complementary to HPGe screening are various forms of mass spectrometry.  Inductively Coupled Plasma Mass Spectrometry (ICP-MS) provides some of the lowest detection limits (sub-ppt, or \SI{0.01}{\micro\becquerel\per\kilogram})~\cite{PNNLcopperICPMS, PNNLpolymerICPMS, JUNO2021acrylicICPMS} available for $^{232}$Th and $^{238}$U as well as other isotopes of interest to the low-background community~\cite{DobsonICPMS,Ra226ICPMS}. 
While ICP-MS can also detect $^{40}$K, interference effects with Ar species produced in the Ar plasma tend to reduce its sensitivity, with ppm levels achieved typically and state-of-the-art instrumentation able to achieve ppb to ppt levels~\cite{PNNLKicpms}. 

One advantage of ICP-MS over HPGe detectors is in the measurement speed. Once the sample is prepared, ICP-MS takes minutes to analyze one sample, whereas the HPGe detector may take weeks. Additionally, smaller sample sizes are required with ICP-MS.  
If laser ablation is utilized, ICP-MS can be a location-specific technique, although this mode of operation requires more complicated calibration techniques typically including the development of certified matrix-matched standards~\cite{LA-ICP-MS,LA-ICP-MScalibrationhard}.

A disadvantage of ICP-MS is in the preparation of the sample (if laser ablation is not used). Optimizing a sample preparation technique for each new material can be time-consuming.  Since digestion or ablation are required, the technique is destructive. 

Most of the underground facilities surveyed either have 1--2 ICP-MS systems on site at their  surface facilities, or have relationships with nearby labs for use of their ICP-MS systems. 
Most of these ICP-MS systems are located in cleanroom facilities with dedicated sample-preparation areas. The experiments surveyed either plan to use these systems or have located other systems within their collaborating institutions.

\subsection{Alpha Screening} 
\label{supportAlpha}

Many alpha detectors have negligible backgrounds reduced by operation underground, but backgrounds of the most sensitive detector for $\alpha$ screening, the XIA UltraLo-1800~\cite{xia}, with a  sensitivity to surface $^{210}$Po $<$ \SI{0.1}{\milli\becquerel\per\square\meter}~\cite{BUNKER2020163870} are reduced by operation underground by about a factor of 3~\cite{xinranliuLRT2019XIA}.  Despite this fact, relatively few underground sites (Boulby, Kamioka, PNNL, and Y2L~\cite{Y2Lxia}) have underground XIA detectors;  one will be moved underground at SNOLAB soon.  Most experiments require surface-alpha sensitivity that may be achieved with the XIA, but improved sensitivity is needed by Argo and is important for many experiments wishing to ensure that assembly occurs within the background requirements, rather than resulting in a need to etch or replace materials after assembly.

\subsection{Radon Emanation Assays} 
\label{supportRnEmanation}

As described in~\cite{snowmassCFWP4}, emanation of radon provides an important radioactive background for most underground physics experiments, so screening candidate materials for Rn directly~\cite{RadonEmanationRauHeusser,RadonEmanationZuzel,RadonEmanationThesisLiu} is an important support for such experiments.  Although radon emanation assays do not have improved sensitivity underground, many experimental systems requiring emanation assays are too large and/or fragile to move to an above-ground site for assay, and assaying as-built systems underground may be advantageous (see e.g.~\cite{lz_assay_2020}).  
For these reasons, several underground laboratories, including Kamioka, SNOLAB, Boulby, and Canfranc, have radon emanation systems on-site, while SURF has the capability to harvest radon on-site for measurement nearby at South Dakota Mines~\cite{lz_assay_2020}.  

The amount of radon emanation capacity worldwide appears sufficient for future experiments so long as this capacity may be efficiently exploited.  However, for many experiments, improved radon emanation assay sensitivity would be useful, as many measurements of individual materials at the limit of sensitivity may easily add up to total radon emanation higher than the experiment requirements.  Furthermore, ambiguities in interpretation from radon emanation measurements at room temperature when applied to experiments at low temperatures provide a need for future facilities for radon emanation at low temperatures. 

\section{Other Underground Support Needs} 
\label{SupportOther}

Experiments require additional specialized underground support to allow fabrication and assembly of detectors, or to allow experimental specifications to be met during operation.  These support capabilities include underground storage of materials, on-site (including possibly underground) machining, and glove boxes for even cleaner detector assembly.  These capabilities may require reduced radon environments, as may the detector shielding configurations.

On-site underground fabrication facilities are necessary to prevent cosmogenic activation of completed detector parts.  Such facilities may provide benefit to multiple underground experiments at a site.  Underground electroforming of copper parts can produce $>$$10\times$ lower radioactivity than the cleanest commercially available copper, and so is planned for experiments such as CDEX, NEWS-G, LEGEND, NEXT, and nEXO.~\cite{nEXO2021,MJDassay}
Experiments such as SBD and SuperCDMS would also benefit from electroplating of clean copper onto pre-machined copper pieces~\cite{NEWSG2021CuElectroplating,SuperCDMSsnowmass}.  Underground electroforming capabilities exist at SURF, Canfranc, and PNNL, and facilities are planned for Boulby and SNOLAB.  Additional underground crystal growth and fabrication of Ge detectors (to reduce the cosmogenic production of tritium) would also be beneficial for multiple experiments~\cite{UndergroundGe,SuperCDMSsnowmass,JinPingActivation}, but there are no such facilities currently due to their significant cost.  Several labs (at least SURF, SNOLAB, and Gran Sasso) have underground machine shops.  More extensive underground machine shops for general use would benefit future experiments.  

Most underground sites have plenty of non-cleanroom space available for storage of materials that do not need to be kept in clean conditions.
Such long-term storage is important for letting cosmogenic activation decay away in materials of detectors used for rare-event searches.  Most experiments need only modest storage within cleanroom spaces, with needs captured in the discussion in Sect.~\ref{SupportCleanrooms}.  Some of this storage must be in low-radon volumes in order to reduce radon-daughter plateout onto parts.  Such storage is most easily achieved by bagging materials in radon-impermeable bags or vacuum-tight canisters, and/or placing in gloveboxes or cabinets that are purged with low-radon gas, typically liquid nitrogen boil-off. Radon concentrations at or below 0.1\,mBq/m$^3$ are achievable with such purges.~\cite{xinranliuLRT2015gases,RadonArSimgen2009} 


Several experiments require plants for water purification and radon removal (from the water), scintillator purification and degassing, or chemical spaces with fume hoods.  SNOLAB in particular has excellent facilities for such liquid material purification. 
Finally, several experiments require isotopic purification, with some of these needed to be sited underground, such as Te for SNO+.

\section{Conclusions}
\label{SupportConclusions}


The larger, lower-background experiments planned for the future will require larger support facilities that also enable lower backgrounds than are currently available.  Gaps between existing facilities and future needs include the following:

\begin{itemize}
\item Some experiments require larger and/or cleaner cleanrooms than currently exist.
\item Some experiments require larger and/or lower-radon reduced-radon cleanrooms than currently exist. 
\item Existing surface-screening methods for radon-daughter plate-out are not sufficient to inform experiments during assembly as to whether their needs are met.
\item Most assay needs may be met by existing worldwide capabilities with organized cooperation between facilities and experiments.
\item Improved assay sensitivity is needed for assays of bulk and surface radioactivity for some materials for some experiments, and would be highly beneficial for radon emanation. 
\item More extensive underground machine shops for general use would benefit future experiments.  
\end{itemize}

\pagebreak
\bibliographystyle{JHEP}
\bibliography{references}
\end{document}